\documentclass[aps,twocolumn,showpacs,floatfix, bibnotes]{revtex4}

\usepackage{graphicx}%
\usepackage{dcolumn}
\usepackage{color}
\usepackage{amsmath}
\usepackage{here}

\begin{document}
\begin{sloppy}
\newcommand{\be}{\begin{equation}}
\newcommand{\ee}{\end{equation}}
\newcommand{\bea}{\begin{eqnarray}}
\newcommand{\eea}{\end{eqnarray}}
\newcommand\bibtranslation[1]{English translation: {#1}}
\newcommand\bibfollowup[1]{{#1}}

\newcommand\pictc[5]{\begin{figure}
                       \centerline{
                       \includegraphics[width=#1\columnwidth]{#3}}
                   \protect\caption{\protect\label{fig:#4} #5}
                    \end{figure}            }
\newcommand\pict[4][1.]{\pictc{#1}{!tb}{#2}{#3}{#4}}
\newcommand\rpict[1]{\ref{fig:#1}}

\newcommand\leqt[1]{\protect\label{eq:#1}}
\newcommand\reqtn[1]{\ref{eq:#1}}
\newcommand\reqt[1]{(\reqtn{#1})}

\newcounter{Fig}
\newcommand\pictFig[1]{\pagebreak \centerline{
                   \includegraphics[width=\columnwidth]{#1}}
                   \vspace*{2cm}
                   \centerline{Fig. \protect\addtocounter{Fig}{1}\theFig.}}

\title{Nonlinear magnetoinductive waves and domain walls in composite metamaterials}

\author{Ilya V. Shadrivov$^1$}
\author{Alexander A. Zharov$^{1,2}$}
\author{Nina A. Zharova$^{1,3}$}
\author{Yuri S. Kivshar$^1$}

\affiliation{$^1$Nonlinear Physics Centre, Research School of
Physical Sciences and Engineering, Australian National University,
Canberra ACT 0200, Australia \\
$^2$Institute for Physics of Microstructures, Russian Academy of
Sciences, Nizhny Novgorod 603950, Russia\\
$^3$Institute of Applied Physics, Russian Academy of Sciences,
Nizhny Novgorod 603600, Russia}

\begin{abstract}
We describe novel physics of nonlinear magnetoinductive waves in left-handed composite metamaterials. We derive the coupled equations for describing
the propagation of magnetoinductive waves, and show that in the
nonlinear regime the magnetic response of a metamaterial may
become bistable. We analyze modulational instability of different
nonlinear states, and also demonstrate that nonlinear
metamaterials may support the propagation of domain walls (kinks)
connecting the regions with the positive and negative
magnetization.
\end{abstract}


\maketitle

The past decade observed many advances in the design and
engineering of artificial structures with unique electromagnetic
response, which have  broadened significantly the range of
possible wave phenomena that can be accessed in experiment. In
particular, it has been shown that the composite structures may
allow realizing the materials with simultaneously negative
dielectric permittivity and magnetic permeability, also known as
left-handed media~\cite{Veselago:1967-517:UFN}, the unique
materials because of their surprising and often counterintuitive
electromagnetic properties. The composite metallic structures
consisting of arrays of wires and split-ring-resonators (SRRs)
have been demonstrated to possess left-handed properties in the
microwave frequency range~\cite{Smith:2000-4184:PRL}.

A standard theoretical approach for analyzing the properties of
composite metamaterials is based on the effective medium
approximation for both linear~\cite{Pendry:1999-2075:ITMT, Kostin, Gorkunov} and nonlinear~\cite{Gorkunov_NL, Zharov:2003-37401:PRL} metamaterials. In such approach a microstructured composite is treated as a homogeneous isotropic medium characterized by effective macroscopic parameters. This approximation is justified when the characteristic scale of the wavelength of the electromagnetic field is much larger than the period of the microstructured medium. Moreover, this effective medium approach is based on a simple averaging over the lattice of micro-elements, and usually it does not take into account any internal (or eigen) modes of the structure, which might exist due to interaction between the individual elements via {\em microscopic} electromagnetic fields. Such eigen modes in the material containing arrays of SRRs are referred to as {\em magnetoinductive waves}~\cite{sham}. 

In this paper we study a novel physics of {\em nonlinear composite metamaterials} stipulated by magnetic interactions between SRRs. We consider a cubic lattice of parallel SRRs and demonstrate that the effective coupling between the resonators is highly anisotropic. We derive discrete
coupled equations and describe the properties of linear and nonlinear magnetoinductive waves. We show that in the nonlinear regime the magnetic response of a metamaterial may become bistable, and we analyze modulational instability of different nonlinear states. We demonstrate that nonlinear metamaterials may support the propagation of domain walls (kinks) connecting the regions of positive and negative magnetization, and simulate their dynamics
numerically.

{\em Model.} We consider a three-dimensional cubic lattice of
identical parallel SRRs, as shown in Fig.~\rpict{geom}(a). We
assume that the SRR slits are infilled with a nonlinear dielectric
with a Kerr-like nonlinear response, i.e., with dielectric
permittivity $\epsilon = \epsilon_l + \alpha|E|^2/|E_c|^2$, where
$E_c$ is the characteristic nonlinear field, and $\alpha = \pm 1$
correspond to the focusing and defocusing nonlinearity,
respectively; the case $\alpha = 0$ describes a linear response.

\pict{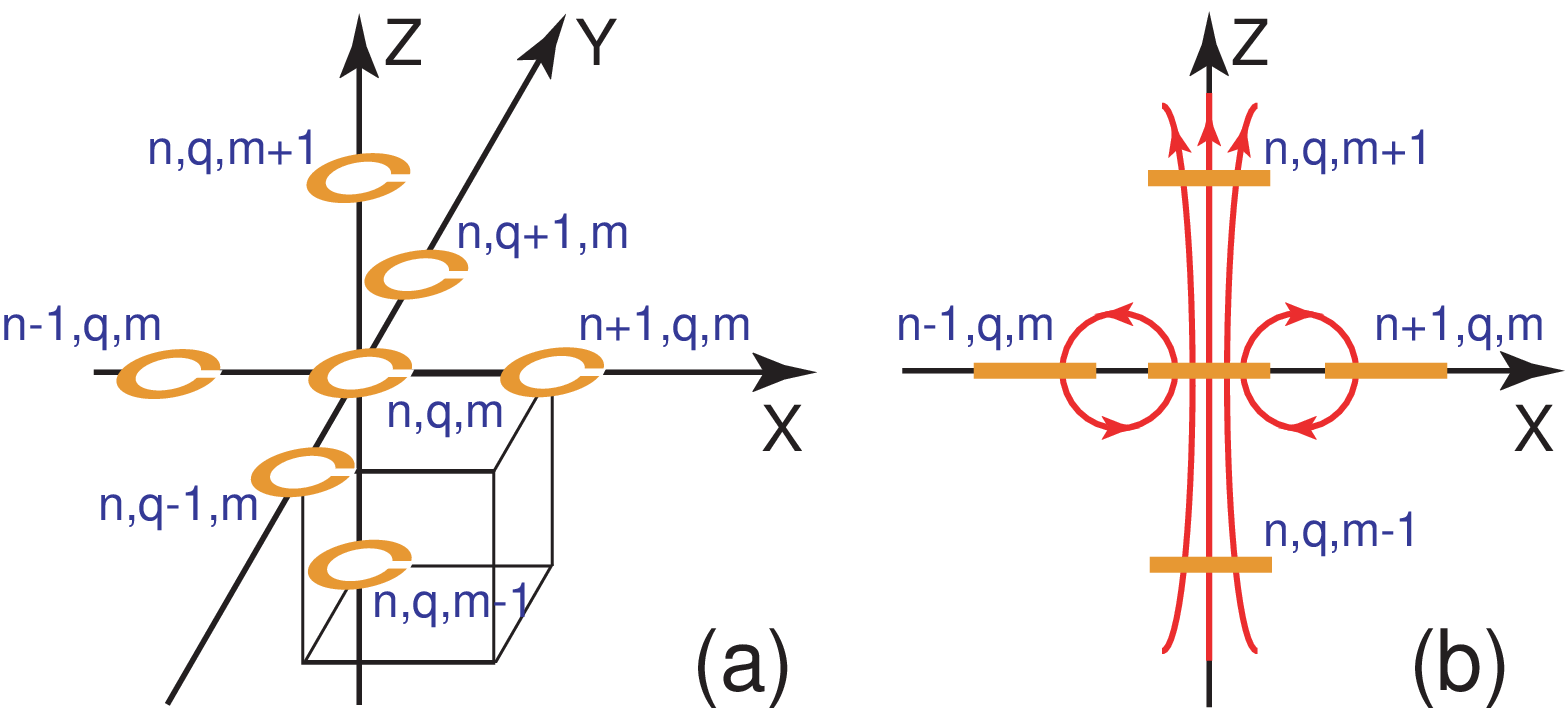}{geom}{(color online) (a) Three-dimensional geometry
and (b) side view of a cubic lattice of split-ring resonators.
Lines in (b) show the direction of microscopic magnetic field.}

An external electromagnetic field induces currents in SRRs; this
generates additional magnetic fields which determine an overall
magnetic response of the composite structure.  In this paper, we take into account the interaction between the nearest neighboring magnetic resonators induced by {\em microscopic fields}, as shown in Fig.~\rpict{geom}(b). This
approximation describes the basic physics due to the SRR
interaction, since the magnetic field of each SRR decays fast
enough and the interaction between the next neighboring magnetic
resonators is weak. In a linear regime, it is possible to take into account interactions between all resonators in the metamaterial sample~\cite{sham}. However, in the nonlinear problem the nearest neighbour approximation simplifies the solution.

We use the indices $n, q, m$ to denote the position of the
resonators along the axis $x, y, z$, respectively. Each SRR
represents an effective oscillatory circuit with inductance $L = 4
\pi a \left[ \ln(8a/r) -7/4 \right]/c^2$ of the loop, nonlinear
capacitance $C_{\rm NL} \approx \epsilon(|E_g|^2) r^2/4 d_g$ of
the slit, and resistance $R=2\pi a/\sigma S_{\rm eff}$ of the SRR
wire (see, e.g., Refs.~\cite{Zharov:2003-37401:PRL,Shadrivov:2004-46615:PRE} and
references therein), where $a$ is the radius of SRR, $r$ is the
radius of the wire, $d_g$ is the size of the SRR slit, $E_g$ is
the electric field induced in the SRR slit, $\sigma$ is the
conductivity of the SRR wire, $S_{\rm eff}$ is the effective
cross-section of the wire, and $c$ is the speed of light.

In the system of magnetically interacting SRRs, we also have an
additional {\em mutual inductance}, described by the matrix
$\hat{M}$, acting between the neighboring resonators. The external
electromotive force ${\cal E}_{n,q,m}$ in each oscillatory circuit
is determined by the external magnetic field $H_0$, ${\cal
E}_{n,q,m} = -(\pi a^2 /c) (dH_0/dt)_{n,q,m}$. Electromotive force
induced in the resonator at the site ($n, q, m$), due to the mutual
inductance with the other resonators, can be written as follows,
\be\leqt{emf_mutual}
e_{n,q,m} = \sum_{n^{\prime}, q^{\prime}, m^{\prime}}
            M_{n^{\prime}-n, q^{\prime}-q, m^{\prime}-m}
            \frac{dI_{n^{\prime}, q^{\prime}, m^{\prime}}}{dt},
\ee
where $M_{n^{\prime}-n, q^{\prime}-q, m^{\prime}-m}$ is the
elements of the mutual inductance matrix describing the
interaction between the resonators at the sites ($n, q, m$) and
($n^{\prime}, q^{\prime}, m^{\prime}$); $M_{0,0,0} = 0$. Using symmetry, we can
present all non-zero elements of the matrix $M$ as follows
\be
M_{1,0,0} = M_{-1,0,0} =M_{0,1,0} =M_{0, -1, 0} = M_{\perp},
\ee
\be M_{0,0,1} = M_{0,0,-1} = - M_{||},
\ee
and derive a set of coupled equations governing the dynamics of
the electric currents in SRRs
\begin{equation} \leqt{volt}
   \begin{array}{l} {\displaystyle
      L \frac{dI_{n,q,m}}{dt} + R I_{n,q,m} + U_{n,q,m} =
    {\cal E}_{n,q,m} +} \\*[9pt]
    {\displaystyle
M_{\perp}
          \frac{d}{dt}[I_{n+1,q,  m} + I_{n-1,q,  m} +
I_{n,  q+1,m} + I_{n,  q-1,m}] -} \\*[9pt] {\displaystyle M_{||}
          \frac{d}{dt} [I_{n,q,m+1} +I_{n,q,m-1}]},
\end{array}
\end{equation}
\be \leqt{curr}
 C_{\rm NL}(|U_{n,q,m}|^2)
\frac{dU_{n,q,m}}{dt} = I_{n,q,m}, \ee
where $U_{n,q,m}$ is the voltage across a slit of the
corresponding resonator. Capacitance of the SRR slit is a sum of
linear and nonlinear parts, $C_{\rm NL}(|U_{n,q,m}|^2) = C_0 +
\Delta C_{\rm NL}( |U_{n,q,m}|^2)$, and we assume that the
nonlinear part is much smaller than the linear one.

We assume that the fields vary harmonically in time [i.e., $\sim
\exp(i \omega t)$] neglecting the generation of higher harmonics,
and use the approximation of the slowly-varying amplitudes for
analyzing Eqs.~\reqt{volt} and \reqt{curr}. The magnetic momentum of
each SRR is proportional to the current in this resonator, ${\bf
m}_{n,q,m} = {\bf z}_0 \pi a^2 {\cal I}_{n,q,m} /2c$, where ${\bf
z}_0$ is the unit vector along the $z$-axis, and ${\cal
I}_{n,q,m}$ is the amplitude of the harmonic current $I_{n,q,m}$.
We are interested mostly in the magnetic response of the structure
near the SRR resonant frequency, $\omega_0 = 1/\sqrt{L C_0}$. Our
choice is motivated by two reasons: (i) nonlinear effects are
essentially enhanced for the frequencies close to the resonant
frequency, and (ii) the left-handed properties of the composite of
wires and SRRs are observed for the frequencies on the right-hand
side from the SRR resonance. Equation~\reqt{volt} gives an
explicit relation between the amplitudes of the current in SRR,
${\cal I}_{n,q,m}$, and the voltage applied across the SRR slit,
${\cal U}_{n,q,m}$:
\be
{\cal U}_{n,q,m} = {\cal I}_{n,q,m}/i \omega
                    \left[
                    C_0 + \Delta C_{NL}\left( |U_g|^2 \right)
                    \right].
\ee
Next, we introduce the dimensionless variables $\tau = \omega_0
t$, $\Omega = (\omega - \omega_0)/\omega_0$, and $\Psi_{n,q,m} =
{\cal I}_{n,q,m}/{\cal I}_c$, where ${\cal I}_c = \omega_0
C_0 U_c$ is the characteristic nonlinear current, $U_c =
E_c \times d_g$ is the characteristic voltage, $\kappa_{||,\perp}
= M_{||,\perp}/L$ are the coupling coefficients, $\gamma =
R/L\omega_0$ is the damping coefficient, and $\Sigma_{n,q,m} = -
\omega H_0 \pi a^2 / c \omega_0 L {\cal I}_c$ is the normalized
electromotive force. The coupling coefficients $\kappa_{||,\perp}$
can be calculated numerically for any SRR geometry, and the
magnetic field of the loop current is well known (see, e.g.,
Ref.~\cite{Jackson:1962:Electrodynamics}). However, here we use an
approximate expression for the magnetic field of the loop current~\cite{Jackson:1962:Electrodynamics},
which yields $\kappa_{\perp} \approx \frac{1}{2}(a/d)^3 = \kappa$
and $\kappa_{||} \approx 2 \kappa$. In the dimensionless units, the
equations for the slowly varying amplitude of the current in SRR
(and, respectively, the magnetic momenta) can be written as
\bea\leqt{psi}
    i \frac{d \Psi_{n,q,m}}{d\tau} -
    \left\{2\Omega - i\gamma + \alpha |\Psi_{n,q,m}|^2
    \right\} \Psi_{n,q,m} -    \Sigma_{n,q,m} \nonumber\\
  =
    2 \kappa
    \left\{
        \Psi_{n,q,m+1} + \Psi_{n,q,m-1} - 2 \Psi_{n,q,m}
    \right\} - \kappa \left\{
        \Psi_{n+1,q,m} + \right. \nonumber\\ \left.
          + \Psi_{n-1,q,m}+ \Psi_{n,q+1,m} + \Psi_{n,q-1,m} - 4 \Psi_{n,q,m}
    \right\},
\eea
where $\tau = \omega_0 t$. 

In addition, Eq.~\reqt{psi} may include long-range effects due to
the interaction with next-neighboring and other SRRs; such effects
will produce additional terms in the right-hand side of
Eq.~\reqt{psi} in the form of higher-order discrete derivatives,
and they can be shown to remain small not changing the major
conclusions of our analysis.  

Equation~\reqt{psi} has a clear physical meaning. The coupling
coefficient, $\kappa$, determines a shift of the oscillator
eigenfrequency due to an effective coupling between SRRs, and the
nonlinear term produces an eigenfrequency shift due to the
nonlinear self-action effect. The effect of the higher-order derivatives will be an additional shift of the resonant frequency. The right-hand side of
Eq.~\reqt{psi} represents the second-order difference operators, which determine the character of the wave propagation and
diffraction. The opposite signs between the respective discrete
difference operators make the derived equations {\em fundamentally
different} from those describing the dynamics of two-dimensional
discrete systems, where the difference operators have the same
sign.

{\em Linear magnetoinductive waves}. First, we describe linear
magnetoinductive waves~\cite{sham} in lossless metamaterials, i.e., we assume
$d/d\tau = 0$, $\alpha = 0$, $\gamma = 0$, and $\Sigma_{n,q,m} =
0$, and substitute $\Psi_{n,q,m} = F(\Omega, {\bf k}) \exp(-i n
k_x d - i q k_y d - i m k_z d)$, where $k_{x,y,z}$ are the
Cartesian components of the wavevector ${\bf k}$. The
corresponding dispersion relation takes the form
\[
\Omega = 4 \kappa \sin^2\left(\frac{k_z
d}{2}\right) -
    2\kappa
        \left[
            \sin^2\left(\frac{k_x d}{2}\right) + \sin^2\left(\frac{k_y d}{2}\right)
        \right],
\]
and its real solutions describe the linear waves with the
frequencies
$|\Omega| < 4\kappa$.
For the dimensional frequencies, this relatively narrow region is
centered at the SRR eigenfrequency. We note that the upper cutoff
of the frequency band of linear waves corresponds to the edge of
the Brillouin zone at $k_x = k_y = 0$ and $k_z = \pi/d$, and in
this limit the linear waves are longitudinal ({\em magnetic
plasmons}). The lower cutoff of the frequency band corresponds to
the point $k_z = 0$, $k_x = k_y = \pi/d$, and the corresponding
magnetoinductive linear waves are transverse. At the resonance
($\Omega = 0$) and in the vicinity of the origin of the Brillouin
zone, the isofrequency surface forms a resonant cone defined as
$2(k_z d)^2 = (k_x d)^2 + (k_y d)^2$, and the phase velocity of
the linear magnetization waves becomes orthogonal to their group
velocity. 

To study the generation and propagation of the magnetoinductive
waves in the SRR lattices, we perform numerical simulations from
Eq.~\reqt{psi}. For simplicity, we consider a two-dimensional
geometry, when all fields are homogeneous along the $y$-axis. We
simulate a lattice of $51 \times 51$ array of SRRs exciting only
one SRR in the middle of the structure (at $n = m = 26$), taking
the damping coefficient $\gamma = 0.01$ and different excitation
frequencies, $\Omega = 0$ and $\Omega = 3\kappa$. To suppress the wave reflection from the boundaries of the numerical domain, we increase smoothly the
damping coefficient in the three rows of resonators near the
boundaries. At the upper cutoff frequency, $\Omega = 3\kappa$,
the longitudinal wave is shown in Fig.~\rpict{lin_waves}(a). In
the case of the resonance, $\Omega = 0$, we observe a resonant
cone as shown in Fig.~\rpict{lin_waves}(b).
\pict{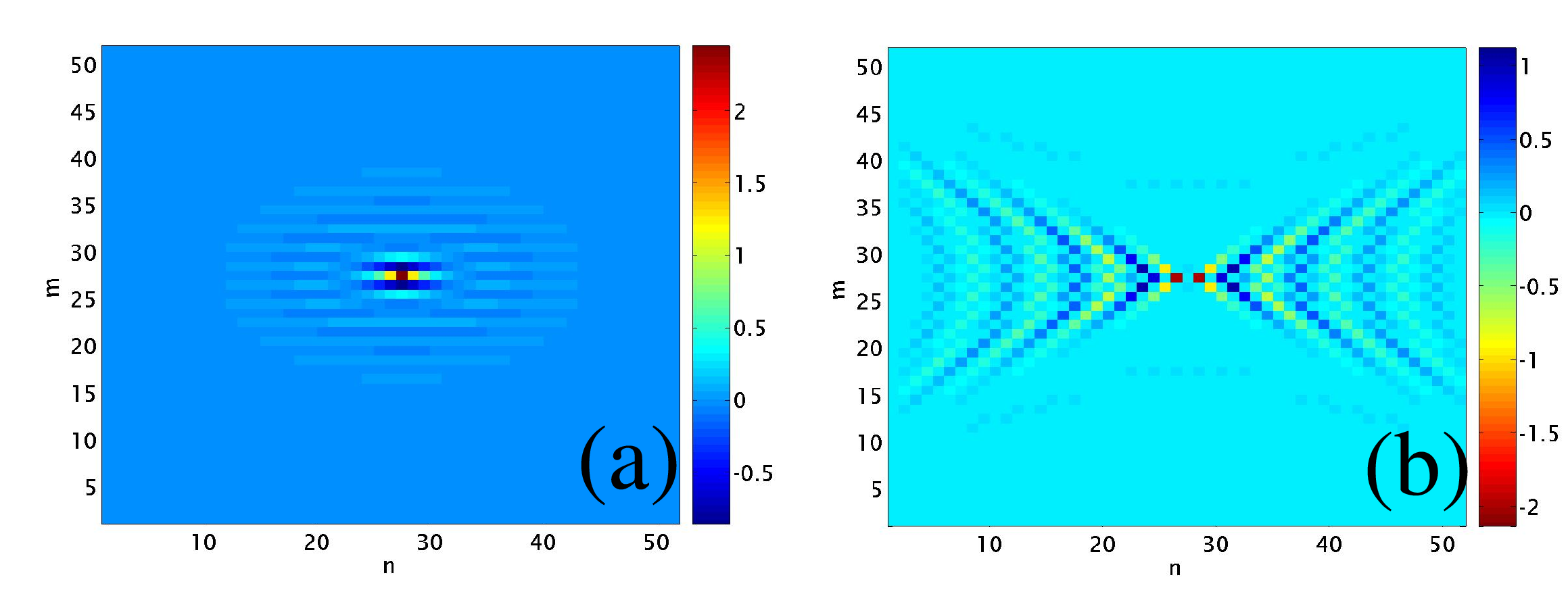}{lin_waves}{(color online) Two-dimensional
anisotropic discrete diffraction of linear magnetoinductive waves
in a lattice of SRRs. Shown is the SRR magnetization on the plane
($n, m$) (a) at the upper cutoff frequency when $\Omega =
3\kappa$, and (b) at the resonant frequency when $\Omega = 0$.}

{\em Nonlinear magnetization.} To study nonlinear effects, we
consider a metamaterial illuminated by a homogeneous
electromagnetic wave. The external electromotive force is the same
in all SRRs: $\Sigma_{n,q,m} = \Sigma_0$. Steady homogeneous
magnetization of the SRR lattice ($\Psi_{n,q,m} = \Psi_0$) can be obtained from Eq.~\reqt{psi} and is described by the nonlinear dispersion, which relates magnetization amplitude with the external magnetic field:
\be\leqt{nonlin_states}
    \left\{
        \left(
            2\Omega + \alpha |\Psi_0|^2
        \right)^2
        + \gamma^2
    \right\} |\Psi_0|^2 =
\left|\Sigma_0 \right|^2. \ee
Two examples of the dependence of $\Psi_0$ on $\Sigma_0$ are shown
in Fig.~\rpict{psi_sigma}. When $\Omega \alpha < 0$, this
dependence becomes {\em multi-valued} [see
Fig.~\rpict{psi_sigma}(a)] and, as is shown below, the middle
branch corresponds to unstable solutions.
Figure~\rpict{psi_sigma}(b) shows the dependence of $\Psi_0$ on
$\Sigma_0$, when $\Omega \alpha > 0$.
\pict{fig03}{psi_sigma}{(Color online) Dependence of the
absolute value of homogeneous magnetization $|\Psi_0|$ of the
metamaterial on the absolute value of the external field
$|\Sigma_0|$ for two cases: (a) $\Omega = 0.2$, $\alpha = -1$,
(b) $\Omega = 0.2$, $\alpha = +1$.}

Such a bistable dependence of the metamaterial parameters may
allow switching between the left-handed transparent and
right-handed opaque states by applying a varying external magnetic
field~\cite{Zharov:2003-37401:PRL}, and may support the existence
of dynamically induced transparent regions inside an opaque
material slab and the formation of spatiotemporal electromagnetic
solitons.

{\em Modulational instability.} Next, we study linear stability of
the homogeneous nonlinear states~\reqt{nonlin_states} with respect
to small perturbations of the form, $\delta \Psi_{n,q,m} = (u + v)
e^{\lambda t} + (u^* - v^*) e^{\lambda^* t}$, where $u = A_1
\exp(ik_x n d + i k_y q d + i k_z m d)$ and $v  = A_2 \exp(-ik_x n
d - i k_y q d - i k_z m d)$. Substituting $\Psi_{n,q,m} = \Psi_0 +
\delta \Psi_{n,q,m}$ into Eqs.~\reqt{psi}, we obtain the following
expression for the instability growth rate $\lambda$,
\be\leqt{lambda}
\lambda = -\gamma + \sqrt{-3|\Psi_0|^4 - 8 \alpha \bar{\Omega} |\Psi_0|^2 -
        4\bar{\Omega}^2},
\ee
where $\bar{\Omega} = \Omega - 2\kappa [ 2\sin^2(k_z d/2) -
\sin^2(k_x d/2) - \sin^2(k_y d/2)]$ can take any value in the range
$[\Omega-4\kappa,\Omega+4\kappa]$. A simple analysis shows that the
real part of $\lambda$ can become positive and, therefore, the
modulational instability can occur when $\alpha \bar{\Omega} < 0$.
In this case, the boundaries of the modulational instability
region are defined from the relations,
\be\leqt{instability} |\Psi_0|^2 = (4\bar{\Omega}/3) \pm
    \sqrt{ (4\bar{\Omega}^2/9) - (\gamma^2/3)}.
\ee
Both the instability and bistability regions can be presented on
the parameter plane ($\Psi_0, \Omega$), as shown in
Fig.~\rpict{psi_omega} for the parameters $\alpha=-1$ and $\kappa
= 0.0025$. The case $\alpha =+1$ can be recovered from the same
results by changing $\Omega \to -\Omega$. The curve that depicts
the instability region consists of three parts. The upper part
represents the boundary \reqt{instability} for $\bar{\Omega} =
\Omega+4\kappa$, which corresponds to the instability with respect
to the excitation of the transverse waves (at $k_z = 0$,
$k_x=k_y=\pi/d$). The lower curve corresponds to the excitation of
the longitudinal waves (at $k_z = \pi/d$, $k_x=k_y=0$). Along the
horizontal line the instability occurs for the waves propagating
at some angle to the main axis of the lattice. Thus, the
instability covers completely the region corresponding to the
middle branch of the bistable dependence shown in
Fig.~\rpict{psi_sigma}(a). Some parts of the upper and lower
branches of that dependence are also unstable. The instability
also occurs for $\alpha\Omega > 0$, as shown in
Fig.~\rpict{psi_omega} (shaded region).
\pict{fig04}{psi_omega}{(Color online) Instability region
(bounded by a solid curve) on the parameter plane ($\Psi_0,
\Omega$). Dashed: the region of the decreasing branch of the
homogeneous magnetization shown in Fig.~\rpict{psi_sigma}(a).
Shaded: instability domain for the magnetization with non-bistable
behavior for $\alpha \Omega > 0$ [see Fig.~\rpict{psi_sigma}(b)].}

Existence of unstable regions can be crucial for engineering
tunable metamaterials with nonlinear properties. The magnetization
waves propagating inside the metamaterial represent alternating
positively and negatively magnetized SRRs. As a result, the
average magnetization of the metamaterial will be dramatically
reduced, possibly suppressing the regions of the negative
effective permeability.
\pict{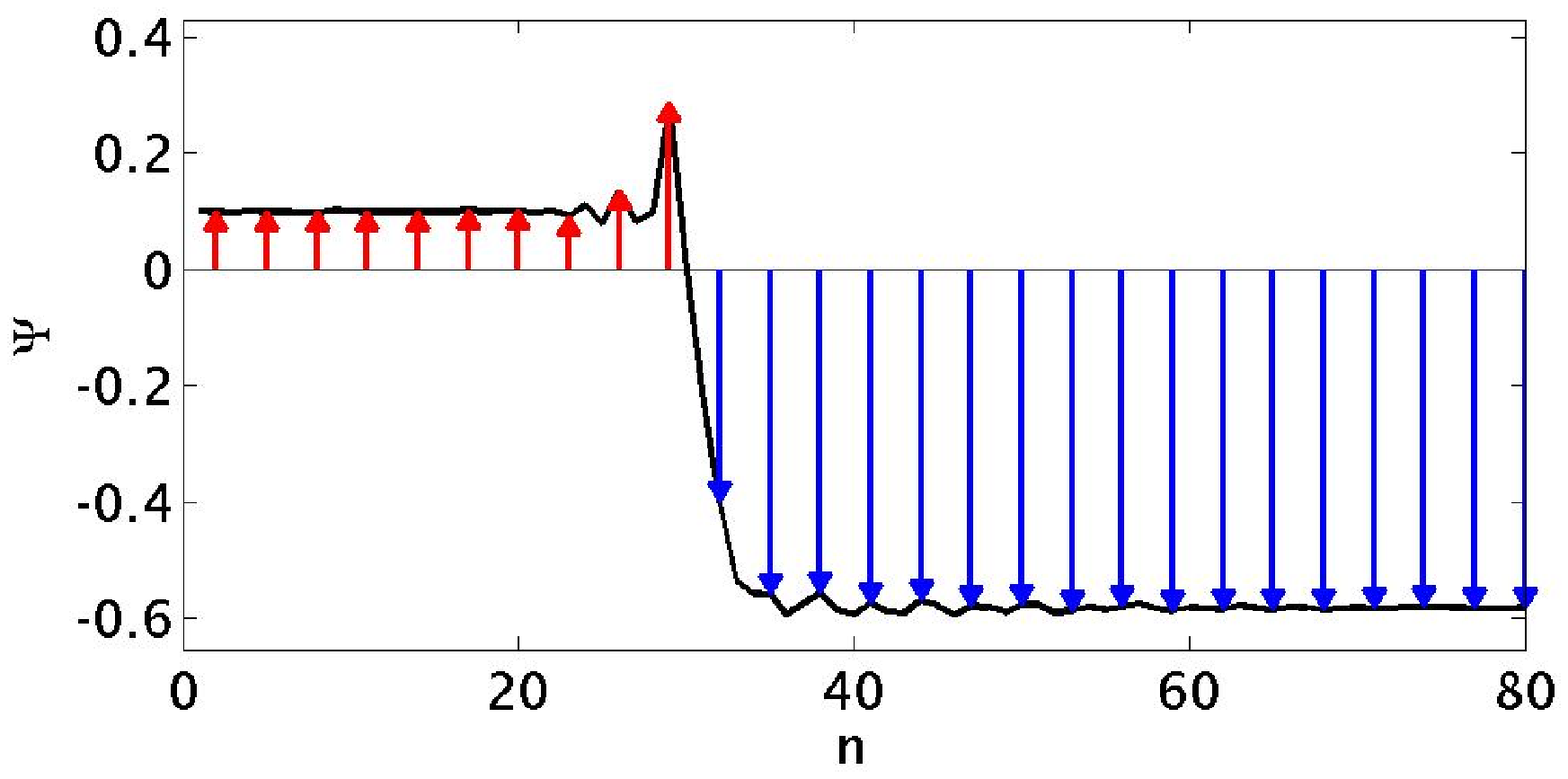}{kink}{(color online) Example of a switching wave of
magnetization (kink) in a nonlinear metamaterial. Shown is the
magnetization $\Psi = \Psi_{n,0,0}$. Arrows show schematically the
magnetization of the corresponding parts of the metamaterial, i.e.,
the magnetization is positive for arrows directed upwards, and
negative, otherwise.}

{\em Domain walls.} Finally, we study numerically nonlinear
switching waves (or domain walls) which can propagate in the
nonlinear metamaterial. We solve Eq.~\reqt{psi} numerically for a
two-dimensional SRR lattice with the size N$\times$N, when $N
=100$, for defocusing nonlinearity ($\alpha = -1$) and $\Omega =
0.2$, when no linear waves exist. First, we create an
inhomogeneous magnetization by a step-like external field
$\Sigma_{n} = \Sigma_0 + \delta \Sigma \times \chi(n-N/2)$, where
$\chi = -1$ for $n<N/2$, and $\chi = +1$, otherwise. We chose
$\Sigma_0$ and $\delta \Sigma$ in such a way that the lower state
excites a homogeneous magnetization of the metamaterial
corresponding to the lower branch of Fig.~\rpict{psi_sigma}, while
the higher state excites the upper branch. After a steady state is
reached, we switch off the inhomogeneous part (put $\delta \Sigma
= 0$), and observe the propagation of a nonlinear switching wave
(kink) with the profile shown in Fig.~\rpict{kink}. Choosing the
parameters $\Sigma_0$ and $\delta \Sigma$, we may control the
speed and direction of the kink propagation. Two out-of-phase
kinks may create a magnetization domain which differs from the
rest of the material. Depending on the external field, such a
domain can collapse, expand, or preserve its shape. The
possibility to control creation and dynamics of such domains seems
to be promising for the design of the structures with controllable
periodic magnetization, photonic crystals, which parameters can be
made tunable.

In conclusion, we have analyzed, for the first time to our
knowledge, nonlinear magnetoinductive waves in composite left-handed
metamaterials. We have derived the effective discrete model that
describes the propagation of magnetoinductive waves in a lattice
of split-ring resonators, and studied both linear and nonlinear
waves, demonstrating they can change dramatically the average
magnetization of left-handed materials. We have revealed that the
bistable magnetic response may lead to the propagation of domain
walls in nonlinear metamaterials.

We acknowledge support from the Australian Research Council and
thank G.V. Shadrivova for help with some figures. AAZ and NAZ
acknowledge a warm hospitality of the Nonlinear Physics Centre.
AAZ acknowledges a financial support from RFBR (grant
N05-02-16357).

\end{sloppy}

\end{document}